# ACPs: Agent Collaboration Protocols for the Internet of Agents

Jun Liu, Ke Yu, Keliang Chen, Ke Li, Yuxinyue Qian, Xiaolian Guo, Haozhe Song, Yinming Li

liujun@bupt.edu.cn

School of Artificial Intelligence, Beijing University of Posts and Telecommunications

*Abstract*—With the rapid advancement of artificial intelligence, the proliferation of autonomous agents has introduced new challenges in interoperability, scalability, and coordination. The Internet of Agents (IoA) aims to interconnect heterogeneous agents through standardized communication protocols, enabling seamless collaboration and intelligent task execution. However, existing agent communication protocols such as MCP, A2A, and ANP remain fragmented and scenario-specific. To address this gap, we propose Agent Collaboration Protocols (ACPs), a comprehensive protocol suite for the IoA. ACPs include registration, discovery, interaction, and tooling protocols to support trustable access, capability orchestration, and workflow construction. We present the architecture, key technologies, and application workflows of ACPs, and demonstrate its effectiveness in a collaborative restaurant booking scenario. ACPs lay the foundation for building a secure, open, and scalable agent interconnecting infrastructure.

## I. INTRODUCTION

In today's digital era, the rapid development of artificial intelligence has given rise to a new type of technical entity—agents. As software or hardware entities endowed with autonomous capabilities for perception, decision-making, and execution, agents have increasingly become a focal point in technological advancement. From basic task automation to complex decision support, agents are being widely applied. They can act as personal assistants to manage schedules and provide information, or as industrial robots to optimize production processes and improve efficiency. With continued technological progress, agents are evolving from rule-based systems to sophisticated entities capable of learning and adapting to their environments. These agents not only accomplish tasks independently but also collaborate with one another to achieve more complex goals, bringing unprecedented convenience to everyday life and business operations.

However, as the number of agents and their application scenarios grow, the limitations of isolated, single-agent systems become increasingly apparent. A single agent has limited ability to handle complex tasks, and lacks sufficient flexibility and adaptability to balance changing environments and multiple goals. Therefore, interoperability and collaboration among diverse agents have emerged as key bottlenecks to be addressed in enhancing their collective capabilities. In response to this challenge, the concept of the Internet of Agents (IoA) has emerged. The IoA is defined as a network formed by interconnecting intelligent agents—capable of autonomous perception, planning, decision-making, and execution—through standardized communication protocols over the Internet. The IoA aims to overcome the siloed nature of individual agent systems by enabling seamless connectivity and efficient cooperation across a distributed network of agents. By dynamically adjusting resource allocation and coordination strategies based on user task requirements, the IoA can significantly boost overall system efficiency and deliver more personalized and intelligent services. For example, a personal assistant agent may cooperate with a hotel booking agent to help users plan a trip, or multiple agents in a smart traffic system may collaborate to reduce congestion and optimize traffic flow. The rise of the IoA could gradually diminish the role of centralized platforms, ushering in a flatter, decentralized Internet where users interact with the digital world via personalized agents that deliver customized services based on individual preferences.

Building the IoA necessitates an interoperable and robust protocol architecture encompassing communication and collaboration protocols. In recent years, notable efforts have been made in this direction. Anthropic introduced the Model Context Protocol (MCP) [1] to facilitate interactions between large models and tools; Google released the Agent2Agent (A2A) protocol [2] for agent communications in enterprise settings; and Chinese independent researcher Gaowei Chang proposed the Agent Network Protocol (ANP) [3] for general-purpose agent interconnection. A comprehensive review of such protocols is provided in the recent ArXiv paper "A Survey of AI Agent Protocols." [4]. Although these protocols have significantly contributed to the development of agent-based systems, they generally focus on limited scenarios. MCP, for example, centers on tool invocation by large language models; A2A aims at enterprise agent connectivity; while ANP adopts a more inclusive vision but lacks provisions for agent access authentication and manageability to preserve open interconnection.

In light of these limitations, and recognizing the IoA as a future-critical infrastructure, this paper proposes a globally-oriented, comprehensive protocol framework called Agent Collaboration Protocols (ACPs). This protocol suite includes protocols for agent registration, discovery, communication, and beyond. It aims to fill the gaps in existing research and lays a standardized foundation for the scalable and reliable evolution of the Internet of Agents.

## II. ARCHITECTURE OF THE INTERNET OF AGENTS

The Internet of Agents (IoA) refers to a network formed by connecting agents—entities capable of autonomous perception, planning, decision-making, and execution—via standardized communication protocols over the Internet. To support its implementation, a typical architecture adopts a five-layer structure, as illustrated in Fig. 1.

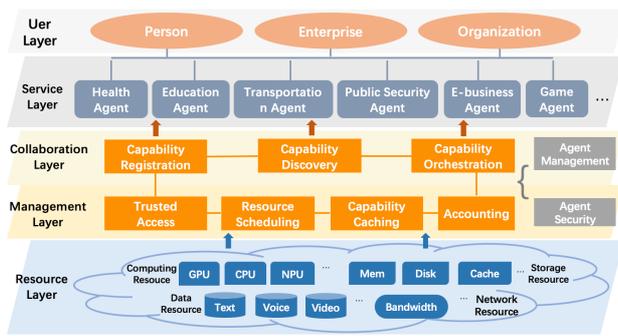

Fig. 1: Architecture of The Internet of Agent

*A. Resource Layer*

The resource layer provides essential computing, storage, communication, and data resources to support the efficient operation and collaboration of agents. Computing resources offer powerful capabilities to support complex algorithms and model execution of agents, including hardware such as CPUs, GPUs, and NPUs, as well as elastic computing services provided by cloud platforms. Storage resources are used to store operational data, model parameters, and user information of agents. These include various storage types such as distributed storage and cloud storage to meet the needs of different agents. Communication resources ensure high-efficiency communication among agents and support various communication protocols and network technologies, including 5G, fixed broadband, Wi-Fi, etc., enabling seamless access and collaboration. Data resources provide rich support for agent operations, including multi-modal data such as text, speech, and video. The resource layer plays a fundamental supporting role in the IoA by not only offering the necessary hardware but also achieving flexible scheduling and efficient utilization of resources through virtualization technologies.

*B. Management Layer*

The management layer is key to ensuring trustworthy collaboration and efficient operation of agents in the IoA. Through a set of mechanisms and architectural designs, the management layer enables agent registration, discovery, task allocation, and collaboration process control, thus building an open, flexible, and efficient collaborative network. Its main capabilities include trusted access, resource scheduling, capability caching, and billing. Trusted access ensures that agents can communicate and collaborate in a secure and reliable environment. Resource scheduling refers to the rational allocation and management of computing, storage, and network resources within the IoA system, enhancing resource utilization and task execution efficiency. Capability caching involves mechanisms and methods to store and reuse agent capabilities to improve system responsiveness and efficiency. Billing defines how to charge and manage user access to agent services.

*C. Collaboration Layer*

The collaboration layer is the core component for enabling efficient collaboration among agents. Its purpose is to support dynamic capability registration, discovery, and orchestration among agents through various mechanisms and architectural designs, thereby achieving effective resolution of complex tasks. When a new agent joins the IoA, it must register and provide detailed descriptions of its capabilities to the server. These descriptions are stored in the data layer of the server, and other agents can query this information to find suitable collaborators. The capability discovery mechanism ensures that registered agents can be located and invoked based on dynamic task planning. Capability orchestration refers to the coordination and management of multiple agent capabilities to efficiently complete complex tasks.

*D. Service Layer*

The service layer consists of agent entities constructed by participants who join the IoA, including enterprises, organizations, or individuals. These agents may possess various capabilities and can offer services either independently or collaboratively. The service layer is the most innovative layer of the IoA and will gain increasing vitality as the IoA's capabilities continue to develop.

*E. User Layer*

The user layer represents the end-users of the IoA, which may include individuals, businesses, or organizational users.

III. CORE CAPABILITIES OF THE INTERNET OF AGENTS

*A. Trusted Authentication of Agents*

Trusted authentication ensures that when agents access the Internet, their identities and behaviors are verified through secure mechanisms. This process is a crucial part of multi-agent collaboration systems, ensuring that agents operate in a safe and reliable environment. The core aspects include:

- Agent Identity Identification: When an agent accesses the Internet, it should carry a digital identifier. This identifier is derived from and bound to the digital identity or communication identifier of the agent's individual or enterprise owner. The identifier is used for agent communication sessions and for mutual authentication and authorization between agents. It should be generated by trusted providers such as telecom operators based on existing identity schemes, to provide a globally standardized identity and naming system. This ensures that agents from different vendors can securely and reliably access the network from any location, enabling interconnection, communication, and collaboration.

- Agent Identity Authentication: In the IoA, technical means should be available to verify the identity of agents, ensuring their legitimacy and trustworthiness in the system. Unlike multi-agent systems developed and operated by a single vendor, the IoA requires cross-platform authentication, allowing agents from different vendors to validate each other's identities. To reduce the risk of single points of failure, peer-to-peer authentication mechanisms should be supported to improve security and system reliability. The authentication process should also be simple and efficient to reduce communication overhead and improve response speed.

- Session Security Between Agents: As the collaboration among agents to complete a template task involves multiple steps, these are carried out through a series of communication sessions within the IoA. Thus, it is necessary to ensure that communication sessions between agents are secure, complete, and confidential. Unauthorized access, data leakage, and malicious attacks must be prevented. Moreover, agents—unlike traditional Internet nodes— possess a high level of autonomy. Therefore, each agent should also be capable of perceiving security risks, detecting security incidents, identifying threats, and autonomously executing response measures to reduce human intervention and enhance its own security.

*B. Agent Capability Registration and Discovery*

Capability registration and discovery refer to the process of registering and managing agent capabilities on the platform so that they can be invoked when needed to complete specific tasks. This process is essential for managing agents and utilizing them efficiently. Key components include:
- Capability Description: Given the complexity of agent capabilities, a standardized description method and language is required. This should express both external capabilities—such as perception, decision-making, action, and interaction—and internal capabilities such as task planning, long- and short-term memory, and self-evolution. These descriptions enable both collaborative execution and effective agent management.
- Capability Registration: Capability registration involves recording and declaring the abilities of agents for subsequent identification and invocation. Similar to capability description, this process should follow clear and standardized protocols to ensure the agents' capabilities are manageable and controllable. To ensure security, capability registration nodes should be operated and managed by highly trusted organizations such as telecom carriers, and can adopt either centralized or distributed architectures.
- Capability Discovery**:** Capability discovery is the process by which agents use standardized protocols to identify and understand the capabilities of one or more agents. The core is to ensure that agents can clearly understand the perceptual, decision-making, action, interaction, and self-management abilities of others. With accurate discovery mechanisms, users or agents can gain a clear understanding of available capabilities in the network, enabling effective automation and intelligence.

*C. Agent Capability Orchestration*

In the IoA, there are many agents with varied or overlapping capabilities. Capability orchestration refers to the coordination and management of multiple agents' capabilities to accomplish complex tasks. This involves distributing tasks, optimizing communication, and decision-making. The complete process includes:
- Task Decomposition: To enable multi-agent collaboration, tasks must first be decomposed into sub-tasks. This increases execution efficiency and speed. Each sub-task is mapped to agents based on their registered capabilities.
- Capability Matching: After decomposition, the sub-tasks are matched with registered agent capabilities. The system then selects the most suitable agents, considering both functional and non-functional requirements (e.g., quality of service). In some cases, agents may be assigned specific roles and grouped into sub-organizations for collaborative execution.
- Task Routing**:** Based on the nature of each sub-task and the matched agents, the system dynamically routes tasks to appropriate agents. Task routing involves distributing both the execution logic and necessary data.
- Task Management and Monitoring: Throughout execution, task status and results must remain visible and controllable. This includes task distribution, communication mechanisms, decision processes, and execution outcomes. Monitoring can be conducted by a management entity (which itself may be an agent), or through autonomous collaboration and self-regulation among agents.

*D. Agent Resource Scheduling*

After task decomposition and capability matching, the next challenge lies in task execution. As with the traditional Internet, there is often a mismatch between available computing/network resources and those required by agents. The issue is especially acute in IoA, where many agents rely on large AI models requiring substantial computational resources. Therefore, resource scheduling becomes a fundamental capability.

Agent resource scheduling refers to the coordinated management of computing resources (CPU, GPU, NPU, etc.) and network resources by management nodes and autonomous agent decision-making, to improve utilization and execution efficiency. It aims to address supply-demand imbalance, network transmission challenges, and equitable resource access. Key technologies include:
- Task Resource Demand Prediction: Using machine learning or deep learning, build models to predict resource demands based on task characteristics, enabling adaptive scaling Capability Matching: After decomposition, the sub-tasks are matched with registered agent capabilities.
- Perception of Computational and Network Resources: Accurate sensing of CPU, GPU, NPU, memory, disk, and network availability, including location, performance assessment, and status monitoring.
- Computational Routing: Similar to data routing, computational routing directs tasks to the most suitable compute nodes based on real-time assessment of resources and network conditions.

*E. Agent Capability Caching*

In a future where nearly all Internet users have multiple personalized agents, agents must follow users across locations and devices. For optimal performance, agent services should be executed as close to the user as possible. Capability caching is thus essential for responsive and efficient operation. o implement caching, the following technologies are required:
- User Location Awareness: Use base station info, IP addresses, GPS, or Wi-Fi data to determine user location—this is the foundation of intelligent caching.
- Agent Capability Storage: Unlike traditional software, agent capabilities rely on code logic as well as

associated data and tools. These must be stored using dedicated mechanisms at caching nodes.

- Capability Invocation Prediction: Analyze usage statistics and user habits to predict which capabilities may be invoked, guiding preloading decisions.
- Cache Scheduling: Given the limited capacity of caching nodes, scientific and efficient scheduling algorithms are needed to maximize performance across user groups.

*F. Agent Usage Billing*

As a platform designed to support widespread commercial applications, the IoA must feature accurate and comprehensive agent billing capabilities. Billing refers to how users are charged for accessing agent services. Different scenarios and user groups may require different billing models. Currently, token-based billing is common. However, as the IoA matures, more flexible models—such as pay-per-result, revenue sharing, or effect-based billing—will be needed. Furthermore, unlike existing systems where billing mostly occurs between agents and users, the IoA will also require billing between agents themselves. Thus, an accurate, fair billing system is crucial. In short, the IoA billing system should support diversified pricing strategies to accommodate the needs of various users and organizations. Transparent, flexible billing not only helps manage budgets and costs but also supports scalable and sustainable agent-based ecosystems.

## IV. ACPs: PROTOCOL SUITE DESIGN

To enable the aforementioned capabilities, a standardized and structured collaboration protocol suite is essential. The Agent Collaboration Protocols (ACPs) represent a system-oriented and open protocol family designed to meet this need. ACPs cover multiple core functionalities, including agent registration, discovery, communication, and tool/resource access. The goal is to fill the gaps in existing protocols—which often focus on narrow scenarios—and provide a unified framework for the Internet of Agents (IoA). Specifically, as illustrated in Fig. 2, the ACPs protocol suite consists of Agent Registration Protocol (ARP), Agent Discovery Protocol (ADP), Agent Interaction Protocol (AIP), Agent Tooling Protocol (ATP).

In addition, to ensure manageability of the system, extended protocols are required, such as Agent Authentication/Authorization/Accounting Protocol (A3AP) for secure access, identity verification, and usage-based billing and Agent Management Protocol (AMP) for real-time monitoring, status reporting, exception handling, and governance of agent behavior.

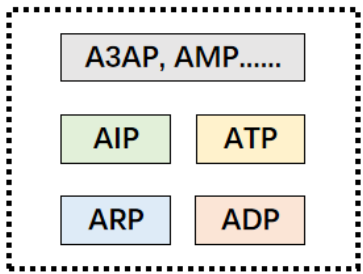

Fig. 2: Protocols in ACPs

### A. Agent Registration Protocol (ARP)

The Agent Registration Protocol (ARP) aims to standardize the declaration, trustworthy registration, and dynamic management of agent capabilities. Its definitions and implementation steps include:

- First, define a capability description language. Based on semantic modeling, this language should support the unified expression of diverse agent capabilities. It must describe both external capabilities (perception, decision-making, action, and interaction) and internal capabilities (task planning, long/short-term memory, self-evolution), enabling not only collaboration but also manageability.
- Second, design a hierarchical architecture of registration servers. The root server manages lower-level registration servers, forming a tree-like structure. Each registration server is responsible for managing the capability data of a group of agents.
- Third, standardize the capability registration process, including identity verification, metadata submission, compliance review, and optional block-chain anchoring.

The interaction process between agents and capability registration servers is illustrated in Fig. 3.

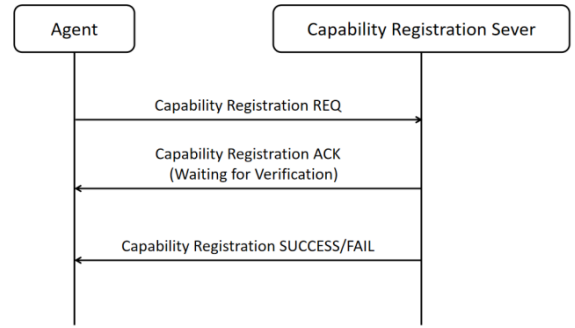

Fig. 3: The Interaction Process Between Agents and Capability Registration Servers

### B. Agent Discovery Protocol (ADP)

The Agent Discovery Protocol (ADP) is designed to enable efficient and accurate capability identification based on task requirements. This allows users or agents to understand the scope and applicability of in-network agents. Key components include:

- First, define a hybrid query language and interface that supports both natural language and structured input for capability queries.
- Second, construct a distributed discovery cloud service. This includes interfaces for effective communication between discovery servers, registration servers, and agents. Capability updates must be propagated efficiently to ensure freshness and consistency.
- Third, implement semantic parsing and alignment mechanisms. This ensures that task requirements (demand side) and agent capability descriptions (supply side) are semantically aligned for accurate matching.

The interaction process between agents and discovery servers is illustrated in Fig. 4.

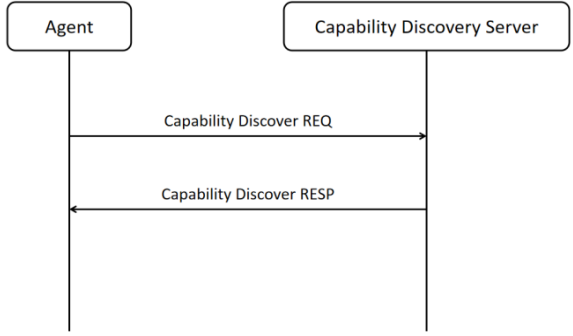

Fig. 4: The Interaction Process Between Agents and Capability Registration Servers

*C. Agent Interaction Protocol (AIP)*

The Agent Interaction Protocol (AIP) supports the construction of task-oriented collaboration frameworks. By enabling dynamic task distribution and intra-group negotiation, AIP ensures reliable execution of complex tasks. Key aspects include:

(1) First, define task assignment mechanisms and workflows. The flow is shown in Fig. 5.

- The user initiates a task request through a personal assistant agent (Personal Agent, P-Agent).
- The P-Agent uses ADP to find agents (Agent_1 to Agent_n) capable of collaborating on the task and sends out group formation requests.
- Each agent responds with a confirmation and joins the task group.
- The P-Agent decomposes the task and orchestrates capabilities, then distributes sub-tasks to each agent in the group.

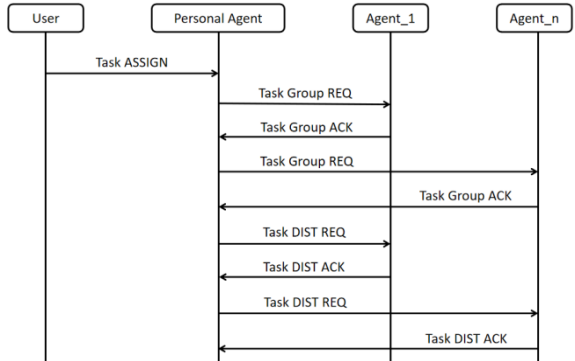

Fig. 5: Task Assignment Mechanism

(2) Second, define task negotiation processes. The flow is shown in Fig. 6.

- Each agent performs its sub-task based on the P-Agent's plan.
- Agents may negotiate with each other or with the P-Agent about task execution details.
- If issues arise during execution, the P-Agent can consult the user for additional inputs.
- Upon completion, results are returned to the P-Agent, which aggregates them and reports to the user.

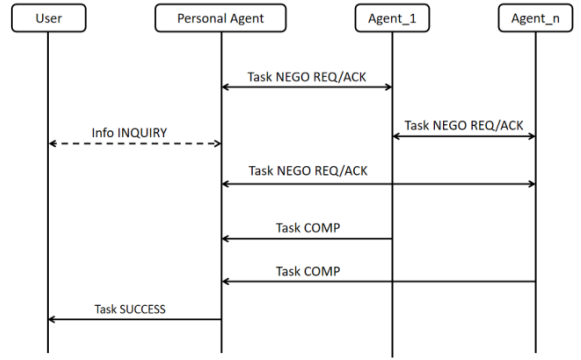

Fig. 6: Task Negotiation Processes

It is worth noting that the A2A and ANP protocols can serve as optional implementation frameworks for the interaction processes defined in AIP.

*D. Agent Tooling Protocol (ATP)*

The Agent Tooling Protocol (ATP) enables agents to dynamically access tools, resources, and external services. It facilitates workflow construction for task execution. Key components include:

(1) First, define standardized methods for accessing tools and resources. The architecture includes:

- Tool Registration Managers: Tools are registered by providers using semantic descriptions and exposed interfaces.
- Resource Managers: Dynamically connect to databases, APIs, and other assets.
- The system supports hierarchical management for scalability.

(2) Second, define context sharing and workflow construction mechanisms. This includes: context managers to maintain state across multi-step tasks; workflow engines to enable orchestration, binding, and monitoring of task flows; execution control mechanisms to ensure high reliability and traceability.

The agent-to-tool/resource interaction process is shown in Fig. 7.

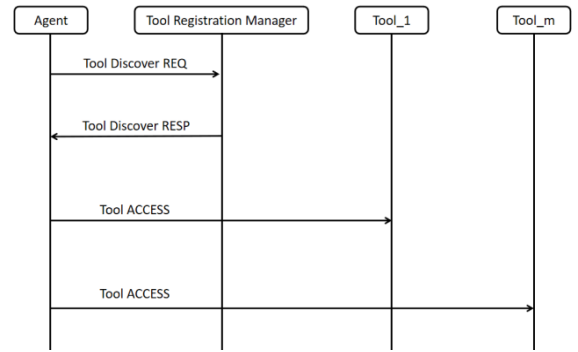

Fig. 7: Agent-to-tool/resource Interaction Process

Note: The Model Context Protocol (MCP) can be considered as a valid implementation framework for ATP.

V. APPLICATION SCENARIOS

To illustrate the application of ACPs, we present a detailed example from the restaurant and travel planning domain. In this scenario, user A is on a business trip and wishes to dine at a local restaurant by 7 PM. The personal assistant agent (P-

Agent) coordinates with multiple other agents through the following steps:

(1) Upon arrival at the destination, user A speaks with the personal assistant agent to describe preferences, including dining time, cuisine type, budget, and specific dietary restrictions (e.g., vegetarian, gluten-free).

(2) The personal assistant authenticates and gains trusted access to the Internet of Agents.

(3) The P-Agent queries the capability discovery server (via ADP) for relevant agents with capabilities in restaurant information search, recommendation, booking, and travel planning, and obtains links to nearby suitable agents.

(4) The P-Agent sends a group formation request to the Restaurant Information Agent, Recommendation Agent, Booking Agent, and Travel Planning Agent, thereby establishing a temporary task collaboration group.

(5) The P-Agent distributes user A's task among the four collaborating agents by assigning them their respective sub-tasks.

(6) Each collaborating agent executes its task by accessing the necessary tools and resources via ATP:

- Information Agent: Searches for restaurants that match the criteria, retrieving data such as name, location, contact, hours, menus, and customer reviews.
- Recommendation Agent: Filters and prioritizes restaurants based on user A's preferences (e.g., ambiance vs. flavor), presenting a ranked shortlist.
- P-Agent: Displays the top choices for user A to review and select.
- Booking Agent: Contacts the selected restaurant to confirm the reservation, including time, number of people, and any special requests.
- Travel Planning Agent: Calculates the optimal route from user A's current location to the restaurant, taking into account traffic conditions and offering to book a ride if necessary.

(7) Upon completion of all sub-tasks, the P-Agent aggregates the responses and reports back to user A with the finalized dining plan and travel details.

To better illustrate the multi-agent collaboration supported by the ACPs protocol family, Fig. 8 presents a layered and distributed architecture of the Internet of Agents for the restaurant reservation scenario. It highlights how different agents—including commander, management, recommendation, booking, information query, booking and travel planning agents—collaborate across core cloud, edge cloud, and access layers.

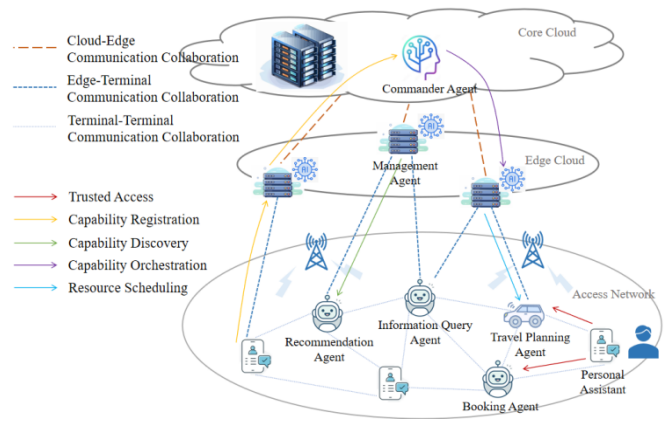

Fig. 8: Multi-agent Collaboration for Restaurant Booking and Travelling Application Scenario

Similarly, ACPs can be applied in other scenarios such as:

- Shopping: The personal assistant agent interacts with search agents and virtual fitting room agents to recommend and simulate clothes try-ons before purchasing.
- Transportation: Agents coordinate flight booking, delay detection, refund processing, and rebooking in real time.
- Home Care and Emergency Response: Agents detect a senior's fall at home, notify family and emergency services, and coordinate real-time communication between medical staff and hospitals.
- Education: An academic assistant agent guides a student through complex problems using multi-modal explanations, and escalates to human tutors if needed.

In each of these, ACPs provide a unified solution framework for seamless multi-agent collaboration. Task flows across domains can be planned clearly and executed reliably, ensuring real-world applicability of the protocol suite.

## VI. CONCLUSION

This paper has introduced the Agent Collaboration Protocols (ACPs), a protocol suite designed specifically for the Internet of Agents (IoA). The framework integrates agent definitions and characteristics, overall architecture, essential system capabilities, and modular protocol design and workflows. Through the core modules—Agent Registration Protocol (ARP), Agent Discovery Protocol (ADP), Agent Interaction Protocol (AIP), and Agent Tooling Protocol (ATP)—the IoA can realize trusted agent access, capability registration and discovery, task collaboration, and dynamic tool/resource invocation.

Using a representative restaurant and travel planning scenario, we demonstrated how these protocols collaborate to satisfy real-world user demands. ACPs enable multi-agent collaboration across domains, with clear orchestration and reliable execution, laying a solid foundation for open, scalable, and intelligent agent ecosystems. The framework offers a unified approach for building next-generation distributed AI systems.

It is important to note that this work presents the first version of the ACPs protocol family. Several aspects, including security mechanisms, message format standards, and performance optimization, remain to be explored further. Future work will focus on refining protocol specifications, designing reference implementations, and validating protocol feasibility in multi-vendor environments. In particular, extended protocols such as A3AP (authentication, authorization, and accounting) and AMP (agent management) require deeper elaboration to ensure real-time governance and cross-agent interoperability.

As research and deployment continue, we believe ACPs can evolve into a core infrastructure for the IoA, offering robust and standardized support for heterogeneous agent collaboration at scale. Ultimately, this will accelerate the development of decentralized, intelligent systems capable of transforming personal life, business operations, and public services.